\documentclass[aps,prl,reprint,groupedaddress]{revtex4-2}

\usepackage{graphicx} 
\usepackage{siunitx}
\usepackage{gensymb}
\usepackage{color}
\usepackage{subcaption}
\usepackage{placeins}
\usepackage{multirow}
\usepackage{tabularx}
\usepackage{xcolor} 

\usepackage[normalem]{ulem} 

\newcommand{\add}[1]{{#1}} 

\newcommand{\movedin}[1]{{#1}} 

\newcommand{\subtract}[1]{} 

\begin{document}



\title{Combining Density-Functional Theory 
with Low-Temperature,
Polarized
Terahertz Spectroscopy
of 
Single Crystals
Explicates the Fundamental Modes 
of 
L-Alanine}

\author{J.~L.~Allen}
\affiliation{Institute for Superconducting and Electronic Materials and School of Physics, 
University of Wollongong, 
Wollongong, 
NSW 2522, 
Australia.}
\email{ja846@uowmail.edu.au}
\author{T.~J.~Sanders}
\affiliation{Institute for Superconducting and Electronic Materials and School of Physics, 
University of Wollongong, 
Wollongong, 
NSW 2522, 
Australia.}
\author{J.~Horvat}
\affiliation{Institute for Superconducting and Electronic Materials and School of Physics, 
University of Wollongong, 
Wollongong, 
NSW 2522, 
Australia.}
\author{K.~C.~Rule}
\affiliation{Australian Centre for Neutron Scattering, 
Australian Nuclear Science and Technology Organisation, 
Lucas Heights, 
NSW 2234, 
Australia}
\author{R.~A.~Lewis}
\affiliation{Institute for Superconducting and Electronic Materials and School of Physics, 
University of Wollongong, 
Wollongong, 
NSW 2522, 
Australia.}


\date{\today}

\begin{abstract}\add{Density-functional theory may be used to
predict both the frequency and the dipole moment of the fundamental oscillations of molecular crystals. Suitably polarized photons at those frequencies excite such 
oscillations.}
\movedin{Thus, in principle, 
terahertz spectroscopy
may confirm the
calculated fundamental modes of amino acids.}
\add{However,}\subtract{While the fundamental modes of amino acids are accessible
in principle by terahertz spectroscopy,}
 reports to date have 
\add{multiple}\subtract{several}
shortcomings:
(a)
material of uncertain purity and morphology and
diluted in a binder material is employed;
(b)
consequently, vibrations along all crystal  axes
are excited simultaneously;
(c)
data is restricted to room temperature,
where resonances are broad and the background dominant;
(d)
comparison with theory has been unsatisfactory
(in part because the theory assumes zero temperature).
Here,
we overcome all four obstacles, 
in reporting detailed
\add{low-temperature}
polarized THz spectra of single-crystal \textsc{l}-alanine,
assigning
vibrational modes using density-functional theory, 
and comparing the calculated dipole moment vector direction
to the electric field polarization of the measured spectra.
\add{Our}\subtract{The} 
direct and detailed comparison of theory with experiment 
\add{corrects}\subtract{has
corrected} previous mode assignments for \textsc{l}-alanine, 
\add{and reveals}\subtract{as well as revealed} 
unreported modes, 
previously obscured by closely-spaced spectral absorptions. 
The fundamental modes are thereby determined.
\end{abstract}


\maketitle



\begin{figure}[t]
	\centering
	\begin{subfigure}{0.5\textwidth}
		\centering
		\includegraphics[width=0.6\textwidth]{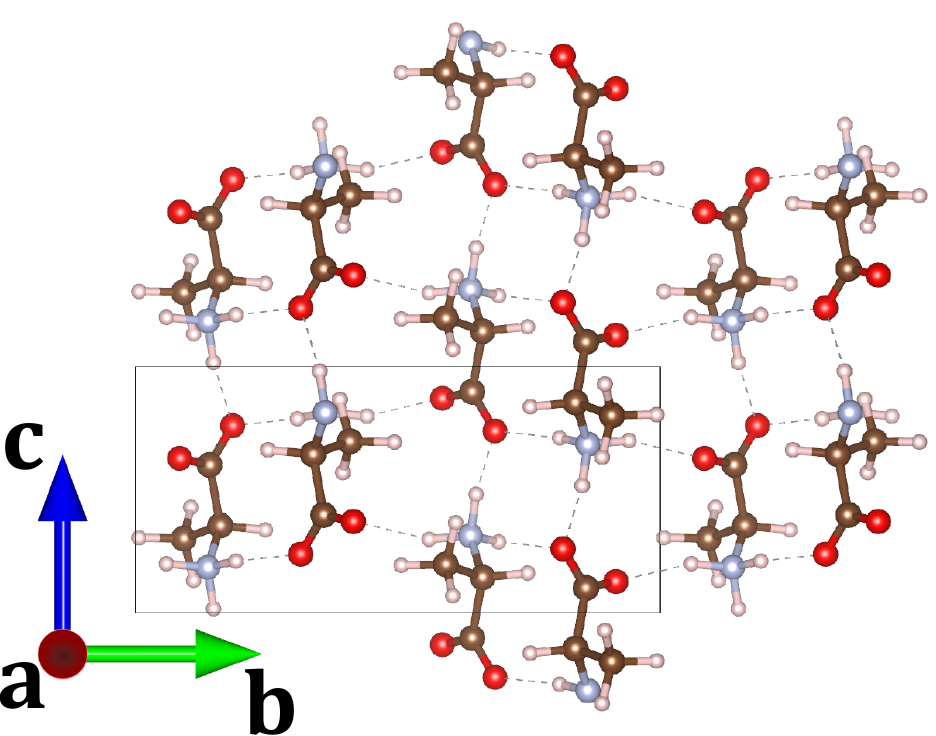}
		\caption{}
		\label{fgr:mol_cell}
	\end{subfigure}
	~
	\centering
	\begin{subfigure}{0.5\textwidth}
		\centering
		\includegraphics[width=0.35\textwidth]{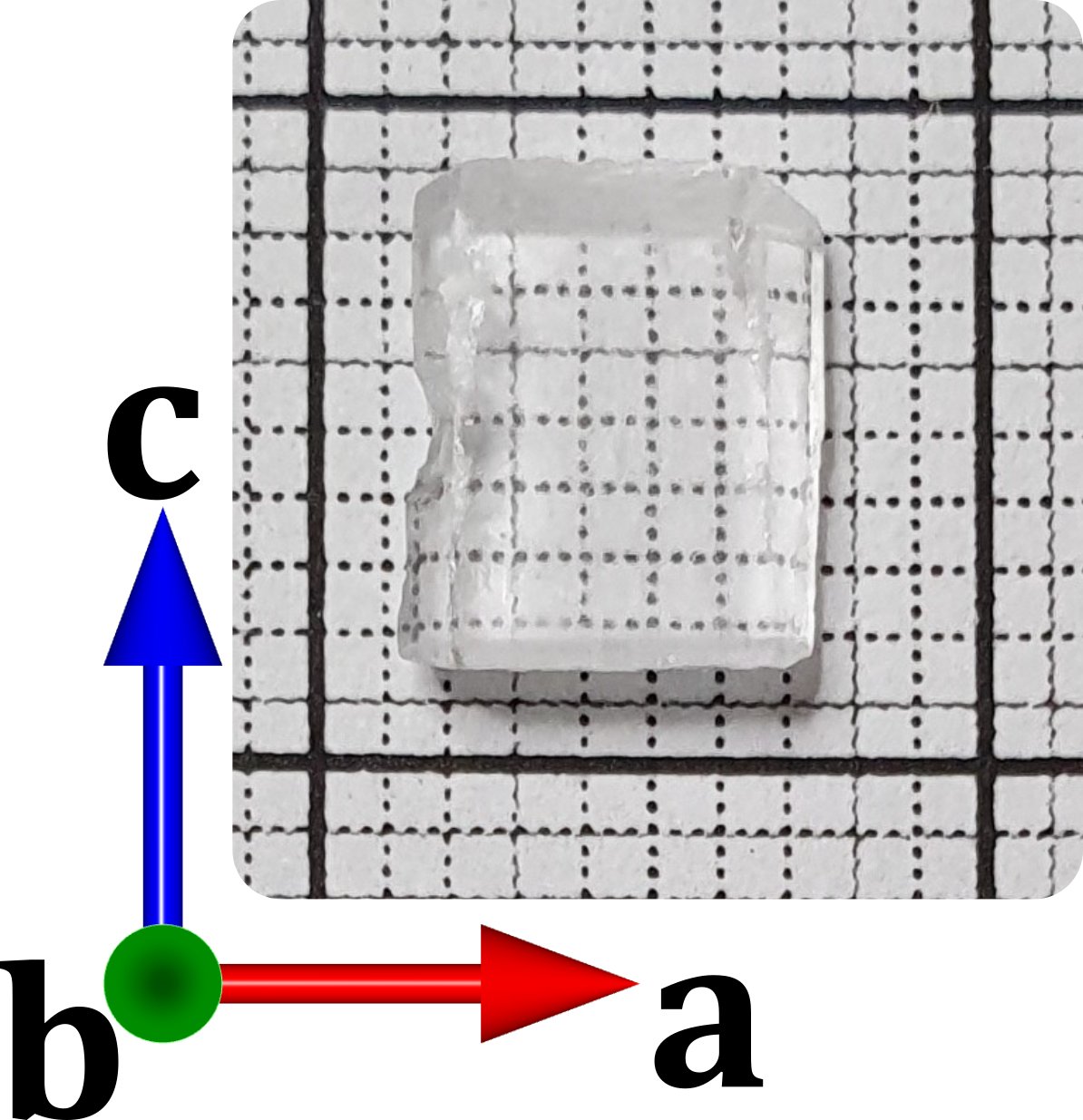}
		\includegraphics[width=0.35\textwidth]{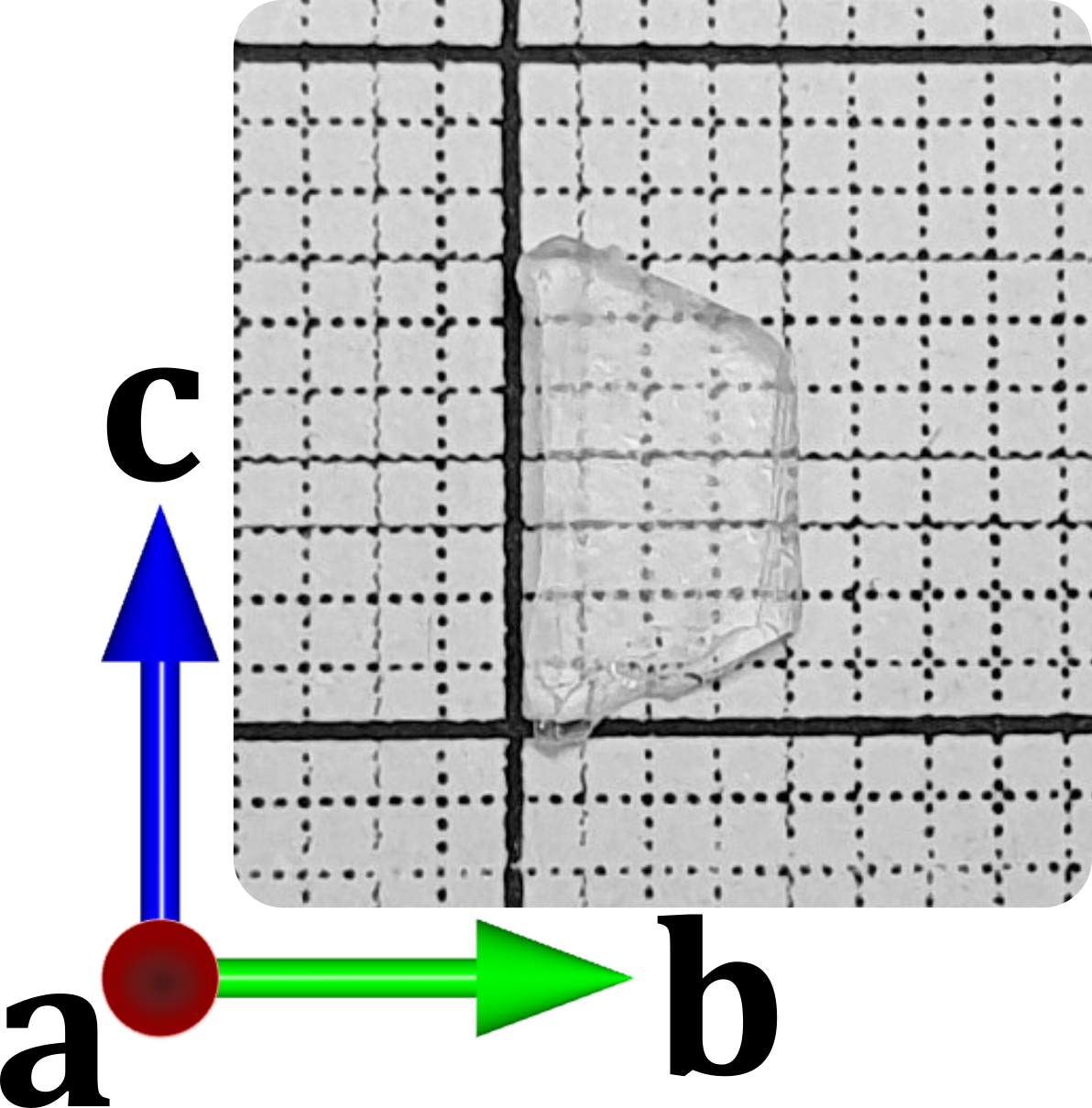}\hfill
		\caption{}
		\label{fgr:samples}
	\end{subfigure}
		~
	\centering
	\begin{subfigure}{0.5\textwidth}
		\centering
		\includegraphics[width=0.6\textwidth]{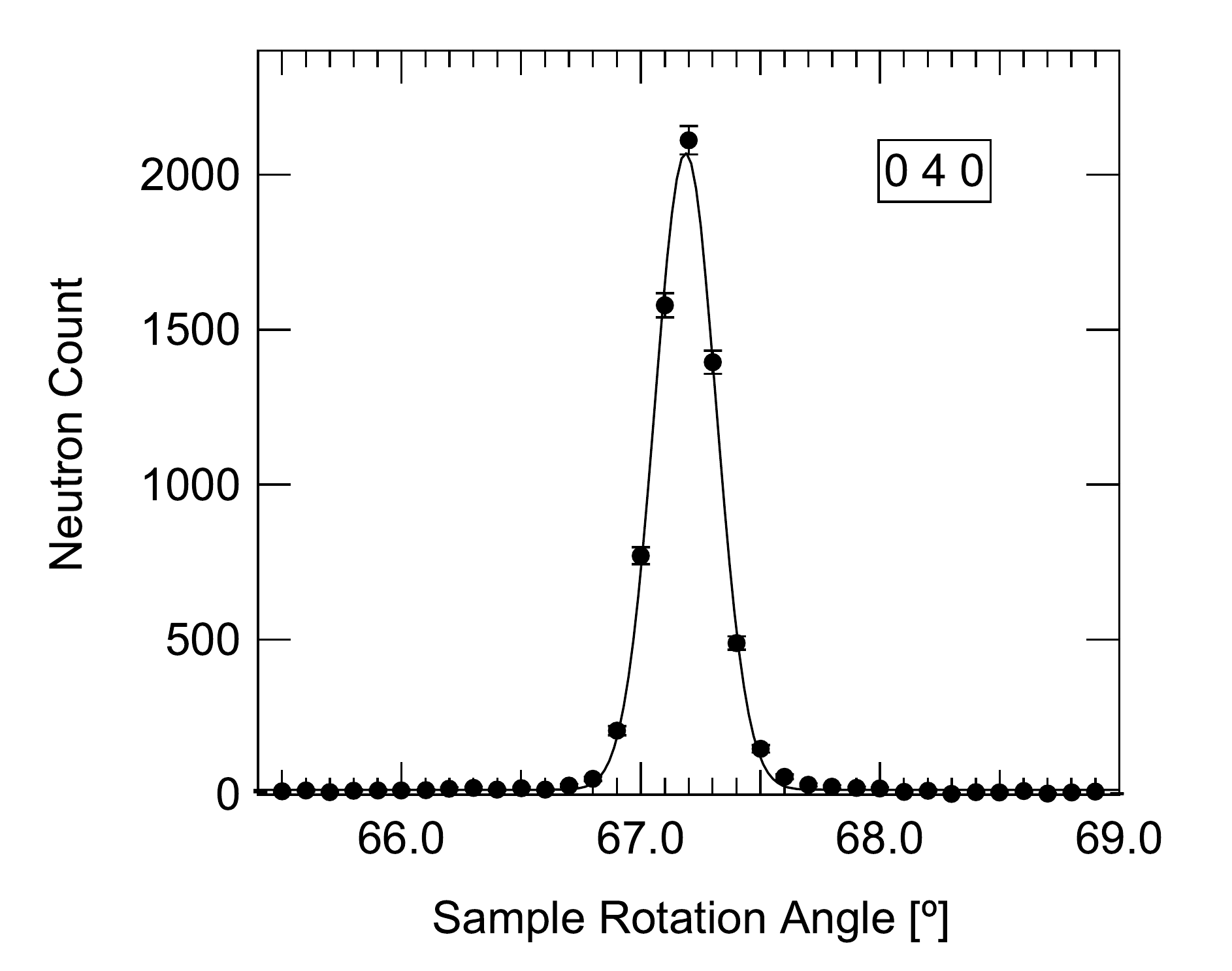}
		\caption{}
    \label{fgr:taipan}
	\end{subfigure}
	\caption{
 (a) 1.5\,$\times$\,2 unit cells of \textsc{l}-alanine. 
 Dashed lines denote hydrogen bonds\subtract{,}
\add{(}with the 
  threshold set to 2.10\,\AA\add{)}. 
 The box represents one unit cell. 
 Structure from 
 Lehmann \textit{et al.} \cite{Lehmann1972}.  
 (b) \textsc{l}-alanine 
 single crystals. 
 On the left is the crystal with $c$-axis vertical
 and $a$-axis horizontal, used for $a$-axis and $c$-axis polarized synchrotron spectra.
 On the right is the crystal with $c$-axis vertical 
 and $b$-axis horizontal,
 used for $b$-axis polarized synchrotron spectra. 
 Both are shown on a 1\,cm-grid. 
 (c) \subtract{The 0\,4\,0 reflection of \textsc{l}-alanine measured at the ANSTO on the Taipan thermal triple-axis spectrometer of a single-crystal sample.} \add{The 
 0\,4\,0 reflection of a single crystal of \textsc{l}-alanine measured with 2.345\,\AA\ neutrons at ANSTO using the Taipan thermal triple-axis spectrometer.
 }
 The solid line is a Gaussian fit to the data. 
 The narrow reflection peak width indicates a high degree of monocrystallinity.}
	\label{fgr:alanine_sample}
\end{figure}

Alanine (C$_3$H$_7$NO$_2$) is an amino acid which is naturally occurring in the human body. It is 
the simplest amino acid to demonstrate chirality. 
Alanine was one of the earliest amino acids, 
fundamental to early life for metabolic processes 
and protein formation \cite{Higgs2009,Kubyshkin2019}. 
In modern life, 
alanine is essential to many peptide and protein structures,
whose 
hydrogen bonds and van der Waals interactions mediate essential biochemical functions. Recently,
the simplest amino acid, glycine, has been found to form in interstellar space 
through a non-energetic mechanism, and it is suspected that the next-simplest amino acid, alanine, 
may 
\add{also}\subtract{undergo a similar process} 
\cite{Ioppolo2020}.
\textsc{l}-Alanine forms a molecular crystal.  Each molecule is in a zwitterionic form, where the amine and carboxylic groups 
\subtract{become }\add{are} 
ionized. 
\movedin{As is
evident in the molecular crystal structure shown in
Fig.\,\ref{fgr:alanine_sample}(a), this
charge distribution}
\subtract{This}gives rise to a dense and complex network
of hydrogen bonds \cite{Lehmann1972}.
\subtract{This is
evident in the molecular crystal structure, as
may be
seen in
Fig.\,\ref{fgr:alanine_sample}(a).} 

Low-energy hydrogen bonds, 
such as those which constitute crystalline \textsc{l}-alanine, 
are associated in energy with electromagnetic radiation in the terahertz (THz) region. 
For this reason,
terahertz spectroscopy has served as an effective probe \add{of}\subtract{into} the intermolecular interactions
of a wide range of biomolecules \cite{Chen2013,ko,shen2007,valine,yi2017,wang2009} and pharmaceuticals \cite{lucia,tetracyline}\subtract{,}
and is an appropriate technique to characterize the interactions in \textsc{l}-alanine. 
Terahertz spectroscopy 
\subtract{can  be }
\add{has been} used to assign vibrational modes,  elucidate the mechanics of protein formation and function \cite{2008He},
and
unravel molecular dynamics  \cite{2020Hutereau}.
\add{More generally,
terahertz physics is an emerging field 
of wide application \cite{tp}.
The terahertz roadmap \cite{roadmap} predicts numerous advances,
which have recently spanned
terahertz axions \cite{axion},
metamaterials \cite{He_2022},
and field-induced ferroelectricity \cite{fif}.}

\add{Returning now 
to molecular crystals, the}\subtract{The} precise origins of
\add{the} terahertz modes 
are 
\add{poorly}\subtract{not often well} understood.
They are usually inferred from the mode energy and crystal structure.
In contrast to the bulk of experimental data reported,
which does not use polarized radiation\subtract{or single crystals},
polarized \add{(`anisotropic')} THz spectroscopy 
\subtract{can be used to
observe }\add{distinguishes} 
vibrational modes with respect to crystal symmetry via the orientation of the dipole moments \cite{Hoshina_2011}.
This technique also permits the observation of modes that would otherwise be obscured by closely
spaced absorption bands \cite{Deng2021,Singh2012}. 
Thus, polarized THz spectroscopy allows for a much more precise
 mode assignment,  yielding better insight into the physical origin of the observed modes.

\add{We now focus on alanine.}
While it has been intensely studied in the low-energy spectral region \cite{MITA2019,Laman,wang2009,LIU2019,DARKWAH2013,Ponseca2010,shen2007,Taulbee2009,Nishizawa2006,Matei,Yamaguchi2005,BARTHES2002},
many questions relating to its molecular dynamics are still to be answered with precision,
\add{due to inadequate theory and experiment}.
This is due to  \subtract{two}\add{five} main factors. 

First\subtract{ly}, density-functional theory (DFT) calculations,
while prolific \cite{Jiang2016,WANG2012,Zheng2012,yi2017,Zhang2015},
in many cases only  agree poorly with experimental spectra. 
The difficulty may be often traced to the use 
of functionals that do not model intermolecular hydrogen bonds well.
\add{As mentioned above, such}\subtract{Such} bonds are largely responsible for
\add{the} low-energy vibrational modes in alanine.
This issue has \add{only} recently been 
\subtract{addressed}\add{identified} \cite{Sanders2021} 
and will be fully 
\add{resolved}\subtract{developed} here.

\subtract{The second limitation to}\add{Secondly, the} precise characterization of \textsc{l}-alanine in the terahertz region
is 
\add{hampered by the lack of} high-quality
\add{samples.}\subtract{experimental data.}
\movedin{Previous studies have commonly used 
blends containing \textsc{l}-alanine in a binding medium for 
pellets \cite{yi2017, LIU2019,Ponseca2010}.
The single-crystal approach of this work 
removes the potential of observing extrinsic features arising from the binder
or interactions
between the binding media and the material under study.}
\add{Moreover, single crystals scatter less light than pellets.}\subtract{By using single crystals}\subtract{of \textsc{l}-alanine,}\subtract{which generate little light scattering.}

\add{Thirdly,}
\movedin{the use of crystals permits the modes associated with different crystallographic axes to be excited by linearly polarized radiation.}
Polarized THz spectroscopy
provides direct information to
compare to DFT modeling
concerning the dipole moment direction.

\add{Fourthly,}\subtract{In addition, }
the polarized data
allows us to separate modes which were previously obscured.
\subtract{
In addition, \add{by} polarized THz spectroscopy 
\add{we have}\subtract{has} been able to separate modes 
which \add{were}\subtract{have} previously \subtract{been} obscured.
Moreover, the polarized data provides
direct information 
to compare to \add{the} DFT modeling by way of dipole moment direction.}
\subtract{Previous studies have commonly used pelletized blends containing \textsc{l}-alanine in a binding medium for samples \cite{yi2017, LIU2019,Ponseca2010}.
The single-crystal approach of this work 
removes the potential of observing extrinsic features arising from the binder
or interactions
between the binding media and the material under study.}
\subtract{Moreover,
the use of crystals permits the modes associated with different crystallographic axes to be excited by linearly polarized radiation.}

\add{Fifthly,} by measuring at low temperatures, where resonances are sharper, 
we have been able to observe previously unidentified modes with unparalleled clarity.

\subtract{Single crystals were grown using the solvent evaporation method \cite{SRINIVASAN2011}. 
The crystals used for measurements are shown in Fig.~\ref{fgr:alanine_sample}(b). 
Two crystals were used,
one with an $a$ face and one with a $b$ face,
to allow all three principal crystallographic axes to be probed.}





\add{We now outline our 
theoretical approach.
Further details are provided in the Supplemental
Material. \cite{supplemental}}
\movedin{DFT}\subtract{Density-functional theory (DFT)} 
 was
used to calculate the vibrational modes in \textsc{l}-alanine.
The CRYSTAL17 package was used \cite{Dovesi2018}. 
Initial
atomic coordinates from Lehmann \textit{et al.}~\cite{Lehmann1972} were the seed for full
geometric optimization of both atomic positions and the unit cell.

Zwitterionic amino-acids have proven very difficult
to model with DFT methods to the accuracy required for obtaining
THz spectra comparable to experiment.
\add{Key to our modeling is the choice 
of an appropriate DFT functional.}
The B97-3c functional was chosen
\add{on three grounds:
it includes an appropriate basis set,
is fast and accurate
in modeling non-covalent interactions that are important for this crystal.}
\movedin{The B97-3c functional}
\subtract{, 
as it }
has been successful in
modeling other molecular crystals \cite{Brandenburg2018,Grimme2010,katsyuba}. 
\add{It uses}\subtract{This
functional} 
\subtract{incorporates  basis set superposition error,}
van der Waals interactions, and short-range basis set corrections. 
Modified
def\add{2}-TZVP basis sets \cite{Weigend2005} are used, 
\subtract{with this functional}
specifically
mTZVP \cite{Weigend2005,Grimme2015}. 
The B97-3c functional is a low-cost Generalized Gradient
Approximation (GGA) functional 2--3 times faster than the 
standard GGA functional when used with the \add{m}\subtract{def2-}TZVP basis
set. 
Even though faster, it has been shown to perform with an
accuracy level better than the standard GGA functionals for light, main
group elements \cite{Brandenburg2018}. 


\add{Since a}\subtract{A}ccurate modeling  of the experimental terahertz
spectrum of \textsc{l}-alanine  has proved to be difficult \cite{yi2017,Jiang2016,Tulip2004}\subtract{.
Hence}, very tight convergence criteria were employed for the
\add{full} geometry optimization.
\add{The details are given in the Supplemental Material \cite{supplemental}.}\subtract{The energy convergence tolerance,
in self-consistent field cycles,
was $7\times10^{-14}$\,hartree/atom 
(2$\times10^{-12}$\,eV/atom). 
The geometrical convergence criteria were: 
maximum energy gradient
$4\times10^{-6}$\,hartree/bohr ($2\times10^{-4}$\,eV/\AA), 
RMS energy gradient
$<$$10^{-6}$\,hartree/bohr ($5\times10^{-5}$\,eV/\AA), 
maximum
atomic displacement $10^{-5}$\,bohr and RMS displacement
$4\times10^{-6}$\,bohr. 
A Monkhorst-Pack grid of 13$\times$13$\times$13 was
used to ensure this high level of convergence. 
The fully converged geometry resulted in a band gap of 5.12\,eV.
This is 8\% above the experimental band gap of 4.75\,eV \cite{Akhtar_2012}.}
The infrared absorption spectrum was calculated  in the harmonic approximation for the final converged
geometry. 
The intensities
of the absorptions were calculated with the Berry phase
approach \cite{Pascale2004,Zicovich-Wilson2004}.

\begin{figure}[t]
	\centering
		\includegraphics[width=0.47\textwidth,trim=0 0 0 0]{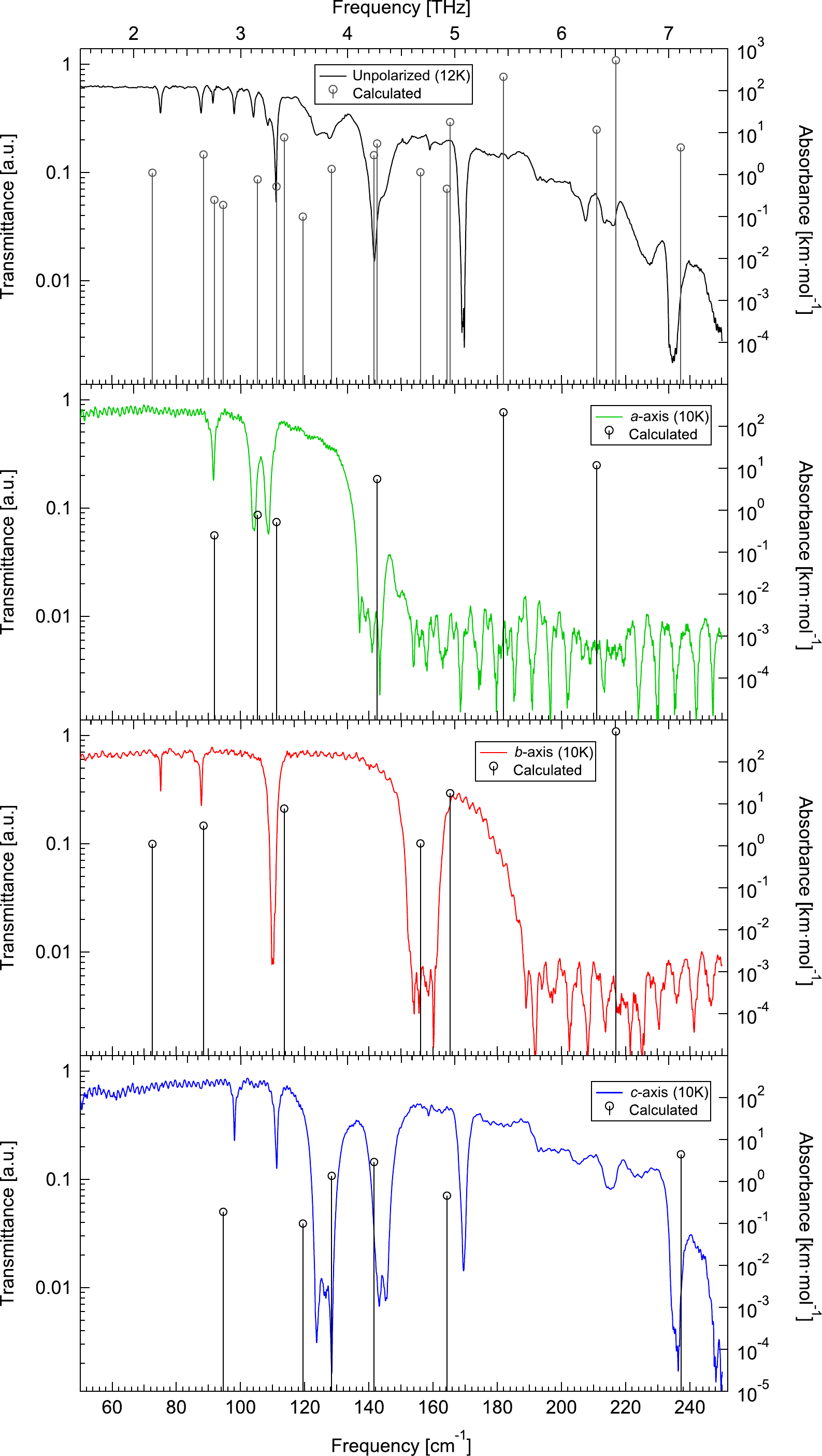}
	\caption{Polarized THz measurements at 10\,K along each crystallographic axis of \textsc{l}-alanine compared to DFT calculation mode positions for the $a$-axis, $b$-axis, and $c$-axis, as well as the unpolarized spectrum at 12\,K.}
	\label{fgr:alanine_pol_singles}
\end{figure}

\begin{figure*}[t]
	\centering
	\begin{subfigure}[!htb]{0.46\textwidth}
		\centering
		\includegraphics[width=1\textwidth, trim={0cm 0cm 0cm 0cm},clip]{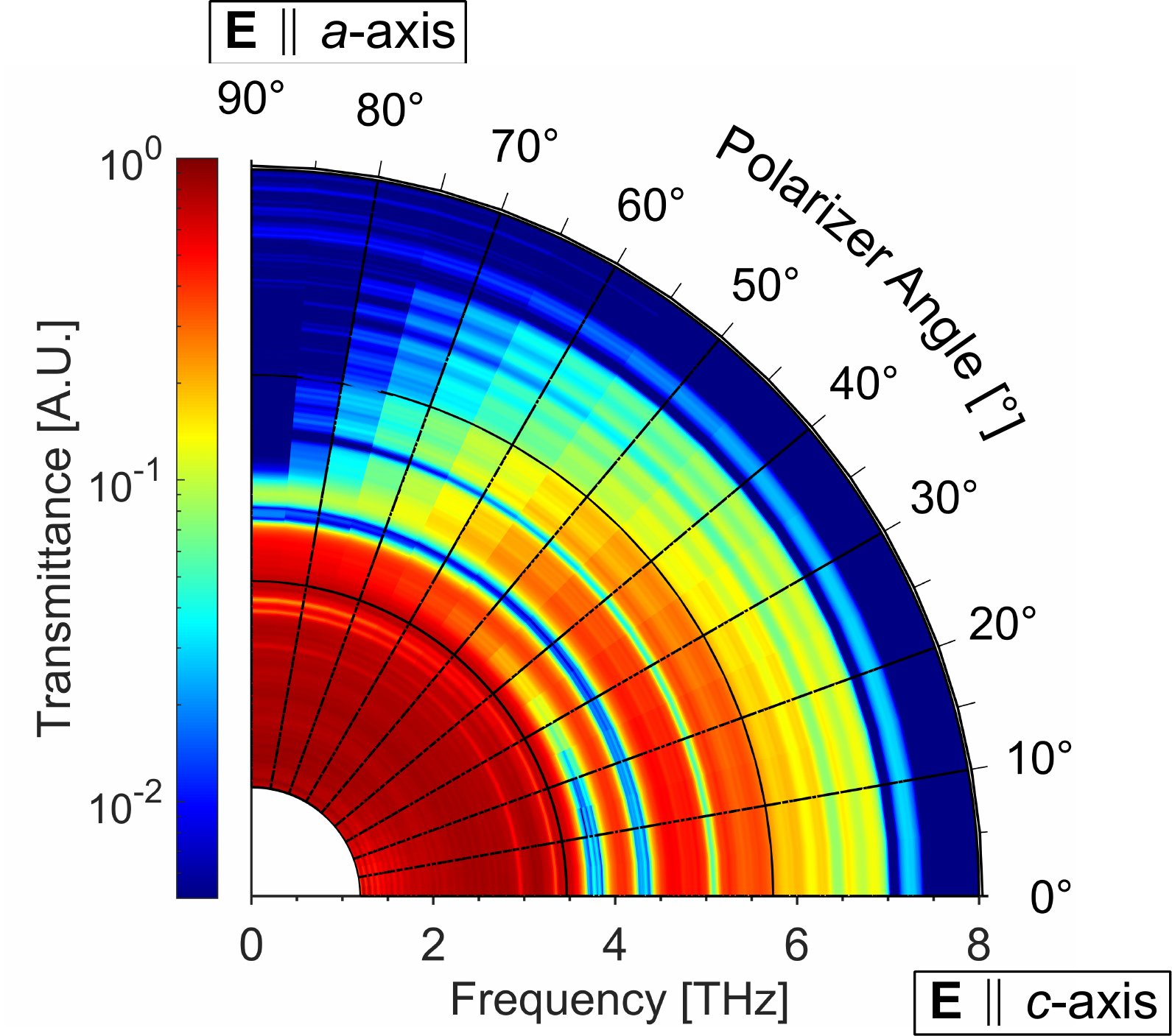}
		\caption{}
		\label{fgr:Radial_B}
	\end{subfigure}%
	~
	\centering
	\begin{subfigure}[!htb]{0.46\textwidth}
		\centering
		\includegraphics[width=1\textwidth, trim={0cm 0cm 0cm 0cm},clip]{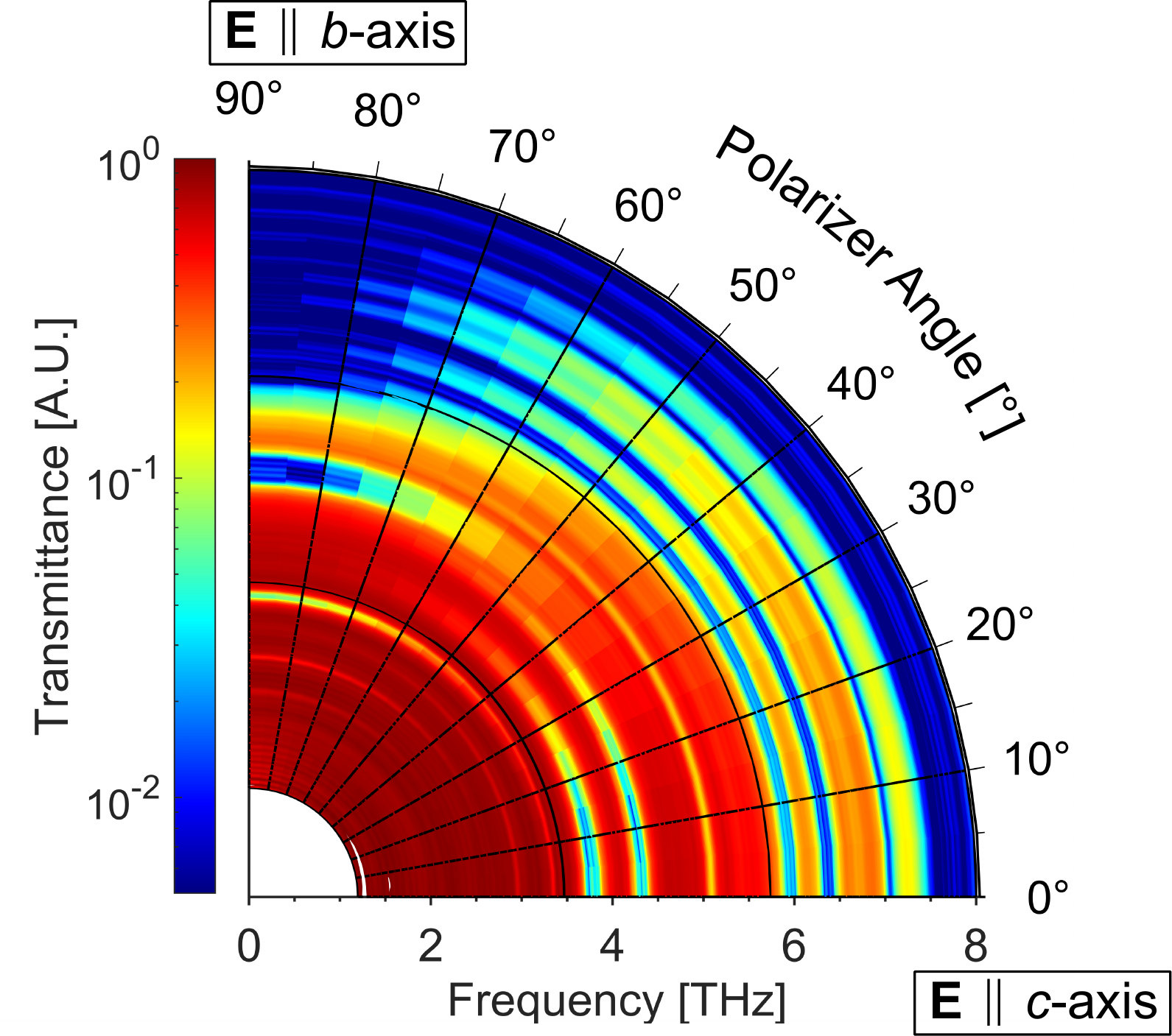}
		\caption{}
		\label{fgr:Radial_A}
	\end{subfigure}
	\caption{Polarized THz measurements 
 at 7\,K \subtract{taken at the Australian Synchrotron}\add{ as the direction of $\vec{E}$ is varied}. 
 (a) $a$-axis to $c$-axis. 
 (b) $b$-axis to $c$-axis.}
	\label{fgr:Radial_plots}
\end{figure*}



\add{We now characterize the samples.}
\movedin{Single crystals were grown using the solvent evaporation method \cite{SRINIVASAN2011}. 
The crystals used for measurements are shown in Fig.~\ref{fgr:alanine_sample}(b). 
Two crystals were used,
one with an $a$\add{-}face and one with a $b$\add{-}face,
to allow all three principal crystallographic axes to be probed.}
\movedin{For optimum anisotropic measurements,  
it is best to ensure that only one crystallographic axis is probed in any one measurement. 
Ideally}
\subtract{,
the incident radiation should  be well-polarized 
and} 
\movedin{the sample should be monocrystalline.} 
\subtract{Polarization of the incident radiation was achieved using a wire-grid polarizer.
This was aligned accurately with respect to a principal axis of the single-crystal samples.} 
\movedin{The morphology of the samples 
was determined using elastic neutron scattering
on the Taipan thermal triple-axis spectrometer \cite{Danilkin2009}
at the Australian Nuclear Science and Technology Organisation (ANSTO),
and agrees with the accepted growth morphology \cite{Massimino_2011}. 
In checking the orientation, 
the 0\,4\,0 reflection peaks in the \textsc{l}-alanine single crystals 
were found to have a low mosaic spread, of $\sim$ 0.8$\degree$, 
as may be seen in the sample rotation scan in Fig.~\ref{fgr:alanine_sample}(c). 
This is (only just) larger than the instrumental resolution of 0.66$\degree$. 
Figure \ref{fgr:alanine_sample}(c) also indicates a good fit of the data to a Gaussian curve, which is expected for a crystalline reflection peak;
moreover, there are no obvious additional peaks. 
This is also true for $\theta$-$2\theta$ scans performed on the crystal samples.
The same holds for the scans done on the 1\,2\,0 reflection peak (as seen in Supplemental Fig.~S3 \cite{supplemental}). 
The good Gaussian fits and the relatively narrow mosaic spreads suggests 
the samples are highly crystalline,
and confirms that they are 
suitable for polarization measurements.}

\add{We now present our results.}
Anisotropic THz spectra of \textsc{l}-alanine 
\subtract{have been}\add{were} measured using synchrotron radiation 
at the Australian Synchrotron.
The light was polarized using a rotatable wire-grid polarizer. 
The polarized experimental transmittance spectra of \textsc{l}-alanine are shown in Fig.~\ref{fgr:alanine_pol_singles}.
The \textit{a}-axis and \textit{c}-axis spectra were measured with the sample having the face perpendicular to the $b$-axis (left-hand sample in Fig.~1(b)).
The \textit{b}-axis spectrum was measured on 
the sample having the face perpendicular to the $a$-axis (right-hand sample in Fig.~1(b)).
The frequency positions and absorbance intensities from \subtract{the}\add{our} \textsc{l}-alanine DFT calculations are also shown on the same plots
for electric dipoles excited by crystalline vibrations along the direction of the each of the three axes.

\begin{table}[b!]
\caption{\ Observed absorption bands positions along each crystallographic axis compared to the closest DFT calculated mode position. Modes are designated by frequency (in THz) with experimental modes at 10\,K. Here, $\parallel$ denotes being parallel with respect to the incident electric field of \add{the} polarized light.}
\begin{tabular}{cccccccc}
\hline
\multicolumn{2}{c}{$\vec{E} \parallel a$-axis} &                      & \multicolumn{2}{c}{$\vec{E} \parallel b$-axis} &  & \multicolumn{2}{c}{$\vec{E} \parallel c$-axis} \\ \cline{1-2} \cline{4-5} \cline{7-8} 
Experiment    & DFT      & \multicolumn{1}{c}{} & Experiment    & DFT      &  & Experiment    & DFT     \\ 
		\hline
		2.74  &  2.76    && 2.25           &   2.18  &&  2.94 &  2.84  \\
		3.12  &  3.16    && 2.63           &   2.65  &&  3.34 &  3.58 \\
		3.26  &  3.33    && 3.30 &   3.41  &&  3.79 &  3.85 \\
		4.25  &  4.28    && 4.71 &   4.68  &&  4.31 &  4.25 \\
		       &  5.46    &&                &   4.96  &&  5.08 &  4.93 \\
		       &  6.33    &&                &   6.51  &&  6.44 &        \\
		       &          &&                &         &&  7.08 &  7.11 \\
		\hline 
	\end{tabular}
	\label{tbl:polarization_modes}
\end{table}

\subtract{For optimum anisotropic measurements,  
it is best to ensure that only one crystallographic axis is probed in any one measurement. 
Ideally,
the incident radiation should  be well-polarized 
and the sample should be monocrystalline. 
Polarization of the incident radiation was achieved using a wire-grid polarizer.
This was aligned accurately with respect to a principal axis of the single-crystal samples. 
The morphology of the samples 
was determined using elastic neutron scattering
on the Taipan thermal triple-axis spectrometer \cite{Danilkin2009}
at the Australian Nuclear Science and Technology Organisation (ANSTO),
and agrees with the accepted growth morphology \cite{Massimino_2011}. 
In checking the orientation, 
the 0\,4\,0 reflection peaks in the \textsc{l}-alanine single crystals 
were found to have a low mosaic spread, of $\sim$ 0.8$\degree$, 
as may be seen in the sample rotation scan in Fig.~\ref{fgr:alanine_sample}(c). 
This is (only just) larger than the instrumental resolution of 0.66$\degree$. 
Figure \ref{fgr:alanine_sample}(c) also indicates a good fit of the data to a Gaussian curve, which is expected for a crystalline reflection peak;
moreover, there are no obvious additional peaks. 
This is also true for $\theta$-$2\theta$ scans performed on the crystal samples.
The same holds for the scans done on the 1\,2\,0 reflection peak (as seen in Supplemental Fig.~S3 \cite{supplemental}). 
The good Gaussian fits and the relatively narrow mosaic spreads suggests 
the samples are highly crystalline,
and confirms that they are 
suitable for polarization measurements.}

The polarized spectra of \textsc{l}-alanine coincide well with the unpolarized spectrum (Fig.~2; also Fig.~S4 and the discussion in Supplemental Material \cite{supplemental}).
All features seen in the unpolarized spectrum appear distinctly in one of the polarized spectra. 
Furthermore, 
calculated modes from DFT also align well with the experimental spectra in Fig.~\ref{fgr:alanine_pol_singles}, 
with the the same number and approximately equivalent positions of the calculated modes and experimental absorption bands. 
Additionally, 
Fig.~\ref{fgr:Radial_plots} shows the evolution of absorption bands as the incident polarization angle is varied. 
All absorption features fade systematically 
as the electric field polarization is rotated away from the axis 
along which the absorption is associated.

With the incident electric field of the THz radiation polarized in the $a$-direction, 
absorption bands are observed at 
2.74, 3.12, and 3.26\,THz,
with a broad absorption band at 
4.25\,THz,
which is not completely resolved as it reaches the noise floor. 
These experimental results are in good agreement with the DFT calculation
when the vibration-induced dipole moment direction is taken into consideration. 
For dipole moments along the $a$-direction, 
DFT calculation predicts vibrational modes at 
2.76, 3.16, 3.33, and 4.28\,THz.
Additional modes are calculated at 
5.46 and 6.33\,THz,
beyond the experimental frequency cutoff of approximately 
4.5\,THz\subtract{,
where the signal to noise ratio drops below 1}.

In the $b$-direction, 
absorption bands are observed at 
2.25, 2.63, and 3.30\,THz.
A broad absorption appears at 
4.71\,THz
and is absorptive enough to reach the noise floor. 
The observations agree with the DFT calculations of modes at 
2.18, 2.65, 3.41, and 4.68\,THz.
A further mode is calculated at 
4.96\,THz.
While not associable with an additional unique experimental absorption beyond that at 
4.71\,THz,
it may be merged with the 
4.71\,THz absorption,
given it is broad and it is incompletely resolved. 
An additional calculated mode at 
6.51\,THz
lies above the cutoff frequency of the $b$-axis polarization at 
5.70\,THz.

The $c$-direction polarization shows absorption bands at 
2.94, 3.34, 3.79, 4.31, 5.08, 6.44, and 7.08\,THz.
DFT calculates modes at 
2.84, 3.58, 3.85, 4.25, 4.93, and 7.11\,THz.
These are all in good agreement with the experimental absorptions, 
with the exception of the observed absorption at 
6.44\,THz.
However, 
this feature does not have the same Lorenzian-like profile as the rest of the absorption bands assigned, 
and could reasonably be a non-resonant feature in origin. 
The frequency cutoff along the $c$-axis is approximately 
7.5\,THz.

Previous work using unpolarized light 
\cite{Sanders2021} assigns the same DFT modes. 
However, it does so to different experimental modes in five cases, 
namely 
the 3.41, 3.58, 4.24, 4.28, and 4.96\,THz modes.
This clearly shows the improved reliability of studies using polarized spectra.
An additional DFT mode at 
7.11\,THz
is identified here
which was not presented in the earlier work.

As mentioned above, 
the polarization data exhibits frequency cutoffs of 
approximately 4.5\,THz
for the $a$-axis, 
5.7\,THz
for the $b$-axis, 
and 
7.5\,THz
for the $c$-axis, 
which is the same for the anisotropic spectra. 
These results are consistent with and extend a previous report on the reflection THz spectrum of \textsc{l}-alanine \cite{MITA2019} 
which shows strong, broad reflection beyond 
4.8\,THz
perpendicular to the $c$-axis, 
but no significant broad, intense reflections parallel to the $c$-axis. 
Thus, it is not absorption, 
but rather strong reflection which most likely affects the frequency cutoff in the $a$- and $b$-directions.

In summary, 
\subtract{the}\add{our DFT} modeling 
\add{of dipoles excited along specific crystal directions}
agrees very well with \add{our}\subtract{the} experimental polarization results.
\add{This}
lend\subtract{ing}\add{s} 
confidence to the physical basis of the DFT calculations\add{.}\subtract{,}
\subtract{Additionally
and s}\add{S}\movedin{pecifically, 
the }\add{success of the} B97-3c functional 
\subtract{used 
in the DFT method}\movedin{in providing very accurate mode calculations} has
\add{verified that weak hydrogen and van der Waals bonds
are critical to the origin of the fundamental modes of
\textsc{l}-alanine.}\subtract{accounting well for the weak hydrogen-bond network in}
\add{Experimentally, t}\subtract{T}hrough the use of polarized THz spectroscopy,
closely-spaced absorption bands have been 
now separated,  
confirming an additional mode at 
3.30\,THz
(with a dipole moment along the $b$-axis).
\add{Comparison with the DFT modelling
has also}
\movedin{extended the assignment of modes made in 
previous work \cite{Sanders2021}.}
\add{Anisotropic theory combined with polarized experiments
thus resolves the fundamental modes of \textsc{l}-alanine.}



This research \add{used}\subtract{was undertaken on} the THz/Far-IR beamline at the Australian Synchrotron, 
part of ANSTO,
the Australian Nuclear Science and Technology Organisation. 
We thank D.~Appadoo and R.~Plathe 
for\subtract{their} assistance. 
Numerical modeling was 
\add{assisted by}\subtract{undertaken 
with the assistance of resources and services 
from} the National Computational Infrastructure (NCI),\subtract{which is} supported by the Australian Government.
We
thank AINSE L\subtract{imi}t\subtract{e}d 
for\subtract{providing} financial assistance (Award-PGRA) to enable this research\subtract{.
This research was},
also supported by\subtract{the} 
Australian Research Council\subtract{(ARC) through 
award} DP160101474.


\bibliographystyle{apsrev4-2}
\bibliography{pol_bib}


\end{document}


\title{Combining Density-Functional Theory 
with Low-Temperature,
Polarized
Terahertz Spectroscopy
of 
Single Crystals
Explicates the Fundamental Modes 
of 
L-Alanine
SUPPLEMENTAL}

\author{J. L. Allen}
\affiliation{Institute for Superconducting and Electronic Materials and School of Physics, 
University of Wollongong, 
Wollongong, 
NSW 2522, 
Australia.}
\email{ja846@uowmail.edu.au}
\author{T. J. Sanders}
\affiliation{Institute for Superconducting and Electronic Materials and School of Physics, 
University of Wollongong, 
Wollongong, 
NSW 2522, 
Australia.}
\author{J. Horvat}
\affiliation{Institute for Superconducting and Electronic Materials and School of Physics, 
University of Wollongong, 
Wollongong, 
NSW 2522, 
Australia.}
\author{K.~C.~Rule}
\affiliation{Australian Centre for Neutron Scattering, 
Australian Nuclear Science and Technology Organisation, 
Lucas Heights, 
NSW 2234, 
Australia}
\author{R.~A.~Lewis}
\affiliation{Institute for Superconducting and Electronic Materials and School of Physics, 
University of Wollongong, 
Wollongong, 
NSW 2522, 
Australia.}


\date{\today}

\maketitle

\add{\section{Further theoretical considerations}}

\add{\subsection{Finer details of the DFT calculation}}

\movedin{
The energy convergence tolerance,
in self-consistent field cycles,
was $7\times10^{-14}$\,hartree/atom 
(2$\times10^{-12}$\,eV/atom). 
The geometrical convergence criteria were: 
maximum energy gradient
$4\times10^{-6}$\,hartree/bohr ($2\times10^{-4}$\,eV/\AA), 
RMS energy gradient
$<$$10^{-6}$\,hartree/bohr ($5\times10^{-5}$\,eV/\AA), 
maximum
atomic displacement $10^{-5}$\,bohr and RMS displacement
$4\times10^{-6}$\,bohr. 
A Monkhorst-Pack grid of 13$\times$13$\times$13 was
used to ensure this high level of convergence. 
\add{
Full geometry optimization was performed, 
including both atomic positions within the unit cell, 
as well as the unit cell parameters.
}
The fully converged geometry resulted in a band gap of 5.12\,eV.
This is 8\% above the experimental band gap of 4.75\,eV \cite{Akhtar_2012}.}

\bigskip\bigskip

\add{
\subsection{Anharmonicity}
}

\add{
In all DFT computational packages,
the normal mode analysis is performed in the harmonic approximation.
In effect,
the calculations assume absolute zero
of thermodynamic temperature;
thermally-induced non-linearities are ignored.
We designed our experiment to match this assumption,
by measuring at low temperatures.}

\add{
At non-zero temperatures,
anharmonic effects come into play.
These may be monitored by tracking the 
change of frequency of the mode with temperature.
Generally speaking,
the modes decrease in energy
(or ``red-shift'')
as temperature increases.
We have previously observed this phenomena for
\textsc{l}-alanine \cite{Sanders2021}
as well as for 
\textsc{dl}-alanine \cite{D0CP05432A},
and for
other amino acids,
including 
\textsc{l}-phenylalanine \cite{ALLEN2021119922},
$\alpha$-glycine \cite{Allen2021},
and 
\textsc{l}-tyrosine \cite{SANDERS2023121970}.
The change in frequency with temperature 
is well-accounted for within a Bose-Einstein model.
}

\add{
In principle,
it may be possible to calculate the true potential along the vector of each normal mode and check by how much each mode deviates from the harmonic potential
as a function of temperature.
However, this is beyond the scope of the present work,
where the main focus is on polarization. 
That is, we calculate the direction of the change of dipolar moment obtained by the modeling  in the harmonic approximation and compare this directly with the measurements taken using light polarized along specific crystal axes at low temperature.}

\newpage
\section{THz instrumentation}

\begin{figure}[h!]
	\centering
		\includegraphics[width=0.9\textwidth]{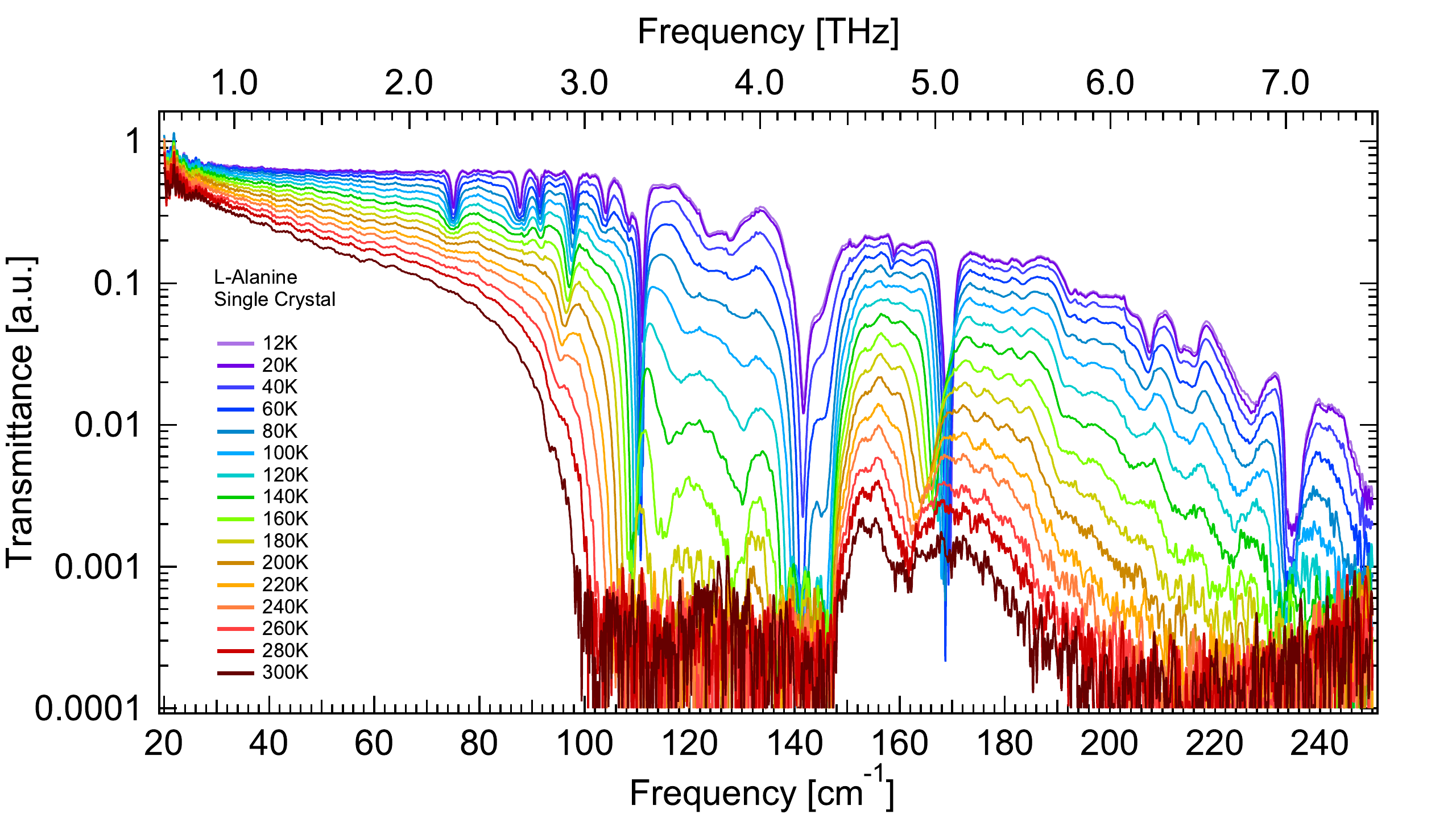}
	\caption{Single-crystal \textsc{l}-alanine spectra measured
 \add{at}\subtract{for} temperatures between 12\,K and 300\,K. Reproduced from Ref.~\cite{Sanders2021}.}
	\label{fgr:alanine_THz_spectra}
\end{figure}

Unpolarized temperature-dependent spectra have been measured on a Fourier transform spectrometer 
(FTS\subtract{: Bomem DA8}) 
to observe modes in the range 1--8\,THz (30--250\,cm$^{-1}$) 
(see Fig.~\ref{fgr:alanine_THz_spectra}). 
A broadband coated mylar beamsplitter was used to measure these spectra at an instrumental resolution of 0.015\,THz (0.5\,cm$^{-1}$) without apodization. 
The Bomem DA8 FTS system was used in conjunction with a THz semiconductor bolometer as the detector. 
A sample temperature range of 12--300\,K was made accessible via a liquid-helium continuous-flow cryostat (Oxford Instruments OptistatCF) 
with an outer vacuum which 
\add{was}\subtract{can be} evacuated to approximately 10$^{-5}$\,mBar. 
The instrumental uncertainty for the temperature was $\pm$0.1\,K. 
Cryostat windows used were polyethylene. 


The absorption bands observed in Fig.~\ref{fgr:alanine_THz_spectra} 
show temperature-dependent frequency shifting. 
This indicates that the absorption bands correspond to vibrational modes, 
which are expected to undergo thermal shifting \cite{Allen2021}.

Polarization measurements were carried out on a separate FTS system (Bruker IFS125) 
with the Australian Synchrotron being used as a bright THz radiation source. 
A 6-$\mu$m multi-layer mylar beamsplitter was used with a 4.2\,K silicon bolometer, 
also at an unapodized resolution of 0.015\,THz. 
A polyethylene (wire) grid polarizer was used to polarize the synchrotron THz radiation incident on the samples.

\newpage
\section{Comparison of c-axis spectra}

\begin{figure}[h!]
	\centering
		\includegraphics[width=0.9\textwidth]{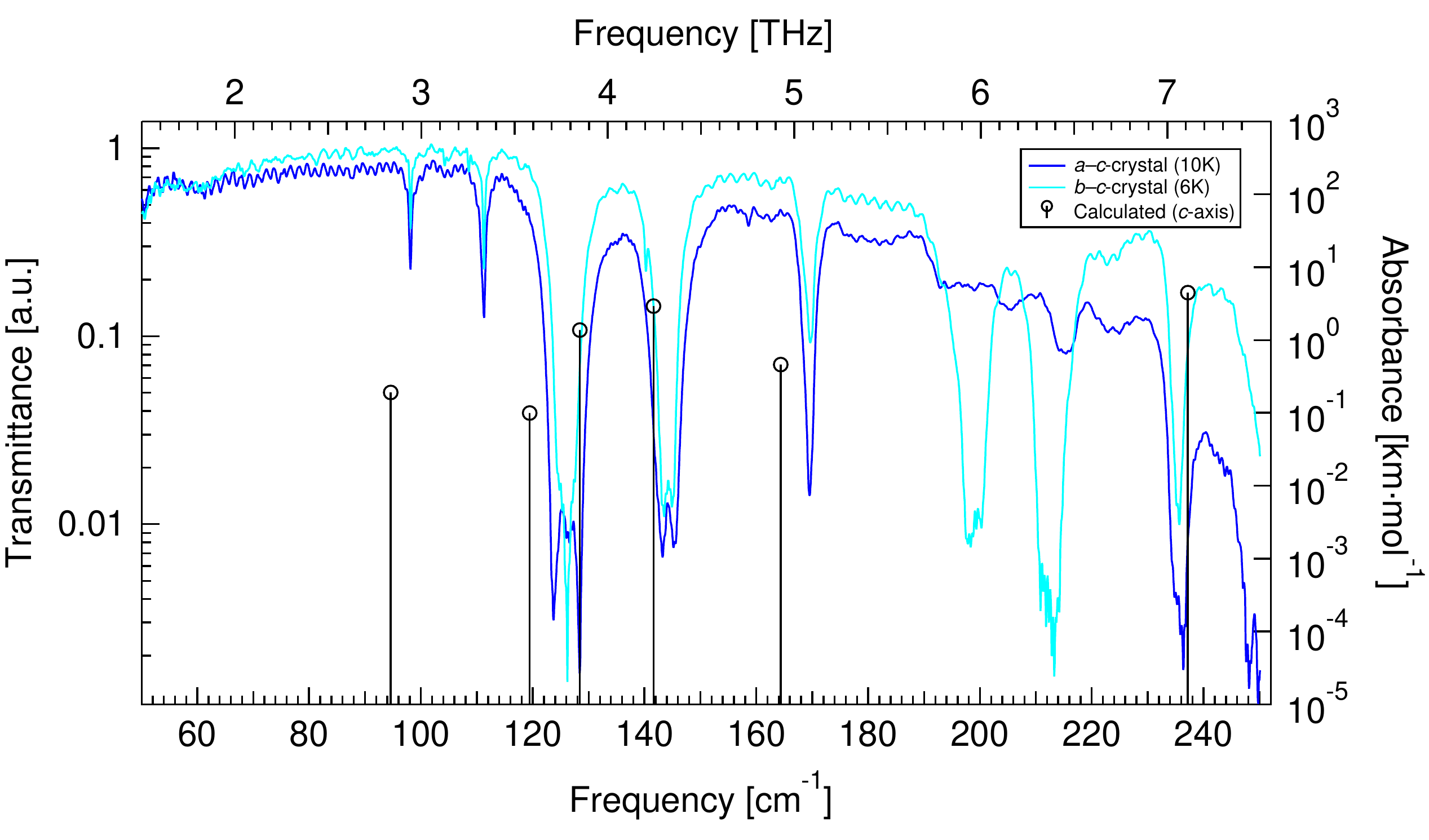}
	\caption{Polarized THz measurements at 10\,K along the crystallographic $c$-axis of \textsc{l}-alanine for both crystals measured. Additional modes appear for the crystal with the $c$--$b$-axes in the normal plane.}
	\label{fgr:c_comparison}
\end{figure}

Spectra for polarization along the $c$-axis were measured using two different crystals, 
since the $c$-axis was the mutual axis for the two crystals used; 
one was sanded to be in the $a$--$c$-plane 
and one in the $b$--$c$-plane. 
Fig.~\ref{fgr:c_comparison} shows the comparison of the $c$-axis spectra for these two crystals. 

The crystal with $b$--$c$-plane geometry shows absorption bands at 
5.85\,THz and 6.33\,THz,
where the alternate crystal in the $a$--$c$-plane geometry does not. 
One may postulate that the origin of these features could be contaminates,
since they only appear in one sample.
However, a more likely explanation for these additional features \subtract{are}\add{is} as components of the $a$-axis modes. While known $a$-axis vibrational modes are not seen at lower frequencies, 
it is usual that higher-frequency absorptions are\subtract{often} stronger. 
So a small component of an axis other than the $c$-axis could have a strong effect at large frequencies. There are DFT calculated modes in the $a$-axis at 
5.43\,THz and 6.30\,THz.
There is only one mode calculated in $b$-axis in this region, 
at 6.51\,THz.
Thus,  these modes can also be reasonably assigned to those calculated in the $a$-axis.

The rest of the spectra for the $c$-axis for each crystal is highly reproducible, which is commensurate with pure samples.

\newpage
\section{Triple-axis neutron scattering}

\begin{figure}[h!]
	\centering
		\includegraphics[width=0.7\textwidth]{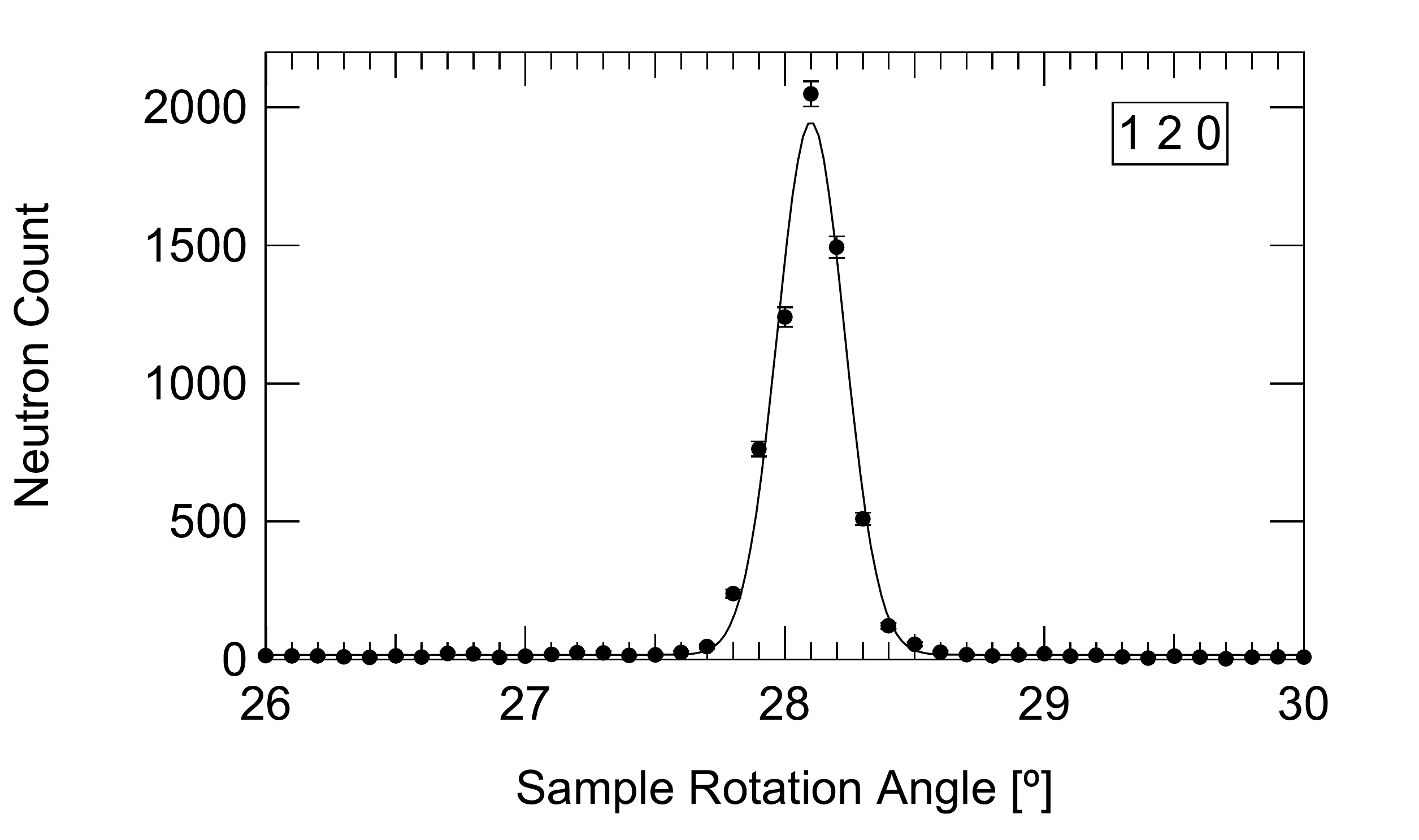}
	\caption{
 \subtract{
 The 1\,2\,0 reflection measured by the triple-axis neutron scattering spectrometer Taipan at ACNS ANSTO. 
 }
 \add{
 The 1\,2\,0 reflection of a single crystal of \textsc{l}-alanine measured with 2.345\,\AA\ neutrons at ANSTO using the Taipan thermal triple-axis spectrometer.
 }
 The 
 \subtract{S1 angle}\add{abscissa}
 refers to the rotation angle of the sample about its center axis.
 }
	\label{fgr:120_reflection}
\end{figure}

The 1\,2\,0 and 0\,4\,0 reflections were measured for the\subtract{grown} \textsc{l}-alanine single crystal
\movedin{grown} \add{with the normal face in the $a$--$c$-plane}. 
Figure \ref{fgr:120_reflection} shows a narrow mosaic spread of $\sim$0.8$\degree$ for the 1\,2\,0 reflection which has been fit\add{ted} with a Gaussian. Additionally, reciprocal space scans show only 0.06\,\add{reciprocal lattice units}\subtract{r.l.u.} as the peak 
\add{full width at half maximum (}FWHM\add{)}. 
This indicates highly ordered single-crystallinity without any twinning effects. 

Neutron scattering probes the entire bulk of a sample\add{,} since uncharged neutrons interact only with the nuclei of atoms\subtract{,} and penetrate into the bulk. 
This corresponds well with the THz spectroscopy measurements which also probe the bulk of the sample. 

\newpage
\section{Comparison to unpolarized spectrum}

\begin{figure}[h!]
	\centering
		\includegraphics[width=0.9\textwidth]{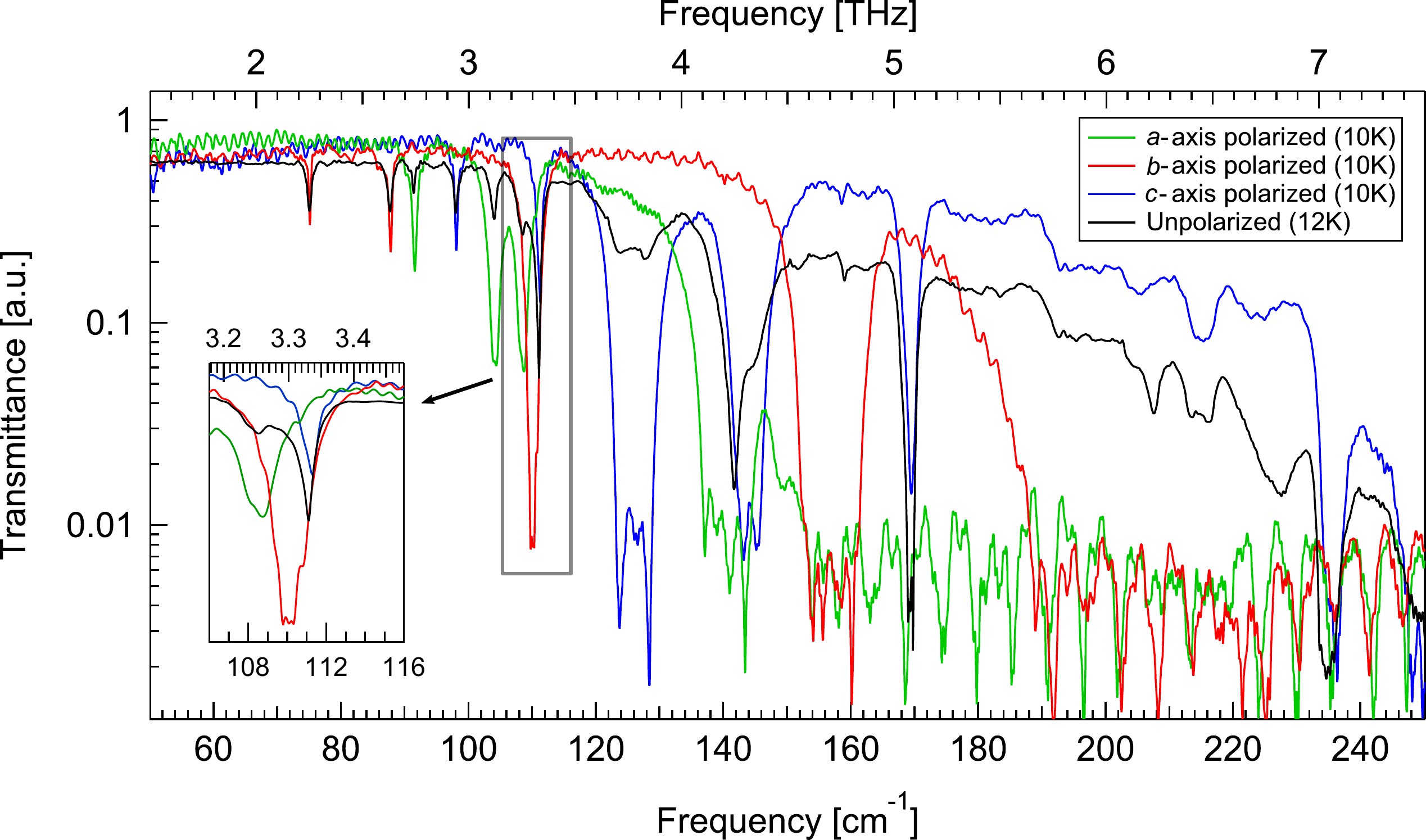}
	\caption{
 Comparison of the polarized spectra with the unpolarized spectrum of single-crys\add{t}alline \textsc{l}-alanine. 
 The inset shows the details at around 3.3\,THz where three vibrational modes overlap. 
 In the unpolarized spectrum, only two peaks are distinguishable.
 }
	\label{fgr:unpol}
\end{figure}

\add{
For ease of comparison,
Fig.~\ref{fgr:unpol}
displays together
the unpolarized spectrum
and the spectra for light polarized along 
each of the three principal crystallographic axes.
We make three observations concerning these.
}

\subtract{The unpolarized spectrum shows all 
absorptions uniquely observed in the \subtract{un}polarized spectra.}

First\subtract{ly}, 
it is clear that the absorption bands in the unpolarized spectrum are all reproduced in one of the polarized spectra.
The anisotropic technique of polarizing the incident beam
\add{successively} along
\add{each of}
the three principal axes of the crystal \add{thus effectively}
works to separate out the vibrational modes associated with electric dipole moments directly along each crystallographic direction. 

\add{Secondly,}
\add{t}\subtract{T}he unpolarized spectrum has no significant contributions from the $a$- nor the $b$-axes above approximately 
5.1\,THz,
since all the light along these directions has been absorbed in this region. 
Thus, only the features corresponding to the $c$-axis contributes to the spectrum in this region. 
The features seen from 
5.7\,THz to 6.9\,THz
in the unpolarized spectrum are the 
non-resonant 
\subtract{modes}\add{features} associated with the $c$-axis.

\add{Thirdly,}
\add{t}\subtract{T}he inset of Fig.~\ref{fgr:unpol} shows the region from 
around 3.3\,THz.
The unpolarized spectrum in this region has only two absorption peaks, at 
3.26\,THz and 3.34\,THz.
These correspond to absorptions in the $a$- and $c$-directions, respectively. 
However\add{,} an extra mode is revealed along the $b$-axis at  
3.30\,THz.
This is obscured in the unpolarized spectrum and has been revealed through the polarized measurements. 
This extends the mode assignment possible with polarized THz spectroscopy. 

\clearpage\newpage
\bibliographystyle{apsrev4-2}
\bibliography{pol_sup_bib.bib}